\documentclass[aps,preprint,groupedaddress,showkeys]{revtex4-2}
\usepackage{float}
\usepackage{multirow}
\usepackage{amsmath}
\usepackage{amssymb}
\usepackage{graphicx}
\usepackage{stackrel}
\usepackage{natbib}
\usepackage[unicode=true]{hyperref}

\begin{document}
	\title{Exploring multimessenger signals from heavy dark matter decay with EDGES 21-cm result and IceCube}
	
	\author{Ashadul Halder}
	\email{ashadul.halder@gmail.com}
	\affiliation{Department of Physics, St. Xavier's College, \\30, Mother Teresa Sarani, Kolkata-700016, India.}
	
	\author{Madhurima Pandey}
	\email{madhurima.pandey@saha.ac.in}
	\affiliation{Theory Division, Saha Institute of Nuclear Physics, HBNI\\1/AF Bidhannagar, Kolkata-700064, India.\\
		Department of Physics, School of Applied Sciences and Humanities, Haldia Institute of Technology, Haldia, West Bengal, 721657, India.}
	
	\author{Debasish Majumdar}
	\email{debasish.majumdar@saha.ac.in}
	\affiliation{Theory Division, Saha Institute of Nuclear Physics, HBNI\\1/AF Bidhannagar, Kolkata-700064, India}
	
	\author{Rupa Basu}
	\email{rupabasu.in@gmail.com}
	\affiliation{Department of Physics, St. Xavier's College, \\30, Mother Teresa Sarani, Kolkata-700016, India}
	
	%\date{\today}
	\begin{abstract}
		The primordial heavy or superheavy dark matter that could be created during the reheating or preheating stage of the Universe can undergo QCD cascade decay process to produce leptons or $\gamma$ as end products. Although these could be rare decays, the energy involved in such decay process can influence 21-cm signal of hyperfine transition of neutral hydrogen during the reionization era. We explore in this work, possible multimessenger signals of such heavy dark matter decays. One of which could be the source of ultra high energy neutrino (of $\sim$ PeV energy regime) signals at IceCube detector whereas the other signal attributes to the cooling/heating of the baryons by the exchange of energy involved in this decay process and its consequent influence on 21-cm signal. The effect of evaporation of primordial black holes and baryon scattering with light cold dark matter are also included in relation to the evolution of the 21-cm signal temperature and their influence are also discussed.
		
	\end{abstract}
	\keywords{21-cm; heavy dark matter; dark matter decay; dark matter - baryon; primordial black hole}
	\pacs{}
	\maketitle

\section{Introduction}
	%Heavy dark matter has turned out to be a topic of interest in high-energy particle physics after the multimessenger observation of IceCube, Fermi-LAT, MAGIC, AGILE, H.E.S.S and HAWC. However, the flairing blazer TXS0506+056 \cite{IceCube:2018dnn} is the only known source, other observed sources are still unknown. The observed muon neutrino flux by IceCube HESE doesnot fit well with single power-law spectrum for $E_{\nu}\gtrapprox2\times 10^5$ GeV (see Fig.~\ref{fig:flx})...........
	 
	The dynamics of the Universe in the dark age are still unexplored due to the lack of luminous sources. The 21-cm neutral hydrogen spectrum can be a promising probe in understanding the dynamics of the early Universe particularly during this unexplored era. The redshifted signature of the 21-cm hydrogen absorption spectrum from the dark ages may provide a detailed understanding regarding the reionization and the Primordial Black Holes (PBHs) \cite{BH_21cm_0,BH_21cm_1,BH_21cm_2,BH_21cm_4,BH_21cm_5,legal_1,Domingo1,Domingo2,Zhou:2021ygz} as well as the baryon-dark matter (DM) scattering and neutrino physics \cite{21cm_nu_1,21cm_nu_1} in the high redshifted epoch.
	
	Keeping this in view, we explore in this work possible multimessenger signals from possible rare decay of fundamental heavy dark matter from dark ages leading to the reionization epoch. Heavy dark matter of mass as high as $10^8$ GeV or more could be created via gravitational production mechanism, non-linear quantum effects during the reheating or preheating stages after inflation. These dark matters are therefore produced non-thermally in the early Universe and are long-lived. Thus these heavy or super heavy dark matter, if exists, exhibits a rare decay process.
	
	In addition, we also consider another thermal WIMP type cold dark matter (CDM) with mass in the range of a few hundred MeV to few GeV (This is consistent with Barkana et~al. \cite{rennan_3GeV}). But the contribution of the heavy dark matter to the total dark matter content of the Universe is assumed to be small and the lighter cold dark matter (CDM) component accounts for the rest of the dark matter content of the Universe. Also, the collisions of the WIMP dark matter with the baryons modify the baryon temperature ($T_b$) and consequently the temperature $T_{21}$ of 21-cm line of neutral hydrogen during the reionization epoch. 
	
	The decay process of primordial super heavy dark matter (HDM) may proceed via QCD cascades. The primary products $q\bar{q}$, on hadronization and decay produce leptons, $\gamma$ etc. as the end products. Ultra high energy (UHE) neutrinos could be one of the final products of this decay process. We attribute these UHE neutrinos from the decay of primordial heavy dark matter to be the source of UHE $\nu$ signal at IceCube neutrino detector in the $\sim$ PeV energy regime. On the other hand, we also explore the energy involved in this HDM decay process and the influence of the consequent heating effect on the evolution of absorption temperature of 21-cm hydrogen line during the reionization era. We have also included other heating/cooling effects that could have been caused by the collisions of cold dark matter particles with baryons and the evaporation of primordial black holes or PBH.
	
	The neutrino events at the IceCube neutrino detector could be an important probe to the sources for UHE neutrinos. The so-called HESE (high energy starting events or contained vertex events) events that are obtained from track and shower events with energy deposition $> 20$ TeV apparently fits a power-law spectrum $\sim E^{-2.9}$. But up-going muon neutron events in the energy range $\sim 120$ TeV and above in PeV energy, show a different power-law dependence. In this work, we consider the IceCube event data points in the latter regime beyond $\sim 120$ GeV and in the PeV region from the 7.5 year IceCube data sample \cite{IC_7.5yr}. We calculate the neutrino flux that can be obtained from heavy dark matter decay of a given mass and then compare the theoretical events with the IceCube experimental data in PeV regime. We may mention here the analysis of similar nature had been addressed earlier by previous authors \cite{nu_line,nu_broad}.

	The ``Experiment to Detect the Global Epoch of Reionization Signature'' (EDGES) \cite{edges} reported a prominent footprint of 21-cm line (21-cm brightness temperature $-500^{+200}_{-500}$ mK) of hydrogen at cosmic dawn ($14<z<20$) with $99\%$ confidence level (C.L.). But following standard cosmology, the estimated brightness temperature can be estimated only $\approxeq-200$ mK. As a consequence, the additional cooling observed by the EDGES experiment can be explained either by enhancing the background temperature $T_{\gamma}$ or by lowering the baryonic temperature which is almost equal to the spin temperature $T_s$ at the epoch $14<z<20$. The temperature of 21-cm absorption spectrum of hydrogen depends on the difference $T_s - T_{\gamma}$ (see later, Eq.~\ref{eq:t21}). The baryon-bark matter interaction is a possible source that cools down the baryonic medium of the cosmos, which induces the larger than the expected difference between $T_s$ and $T_{\gamma}$. Dark matter annihilation, decay, PBH evaporation, dark matter-dark matter interaction could also modify the $(T_s - T_{\gamma})$ \cite{rdi,legal_2,atri,amar21,Mukhopadhyay:2020bml,BH_21cm_0,BH_21cm_1,BH_21cm_2}. In the present work, we address the effects of possible decay of heavy dark matter or HDM on the 21-cm absorption temperature signal. In addition, the influence of PBH evaporation and the baryon-DM interactions are also included in a single framework to study the 21-cm absorption temperature and its evolution.
	%The ``Experiment to Detect the Global Epoch of Reionization Signature'' (EDGES) \cite{edges} reported a prominent footprint of 21-cm line (21-cm brightness temperature $-500^{+200}_{-500}$ mK) of hydrogen at cosmic dawn ($14<z<20$) with $99\%$ confidence level (C.L.). But following standard cosmology, the estimated brightness temperature can be estimated only $\approxeq-200$ mK.
	%As a consequence, the additional cooling observed by the EDGES experiment can be explained either by enhancing the background temperature $T_{\gamma}$ or by lowering the baryonic temperature which is almost equal to the spin temperature at the epoch $14<z<20$. The evaporation of primordial black holes (PBHs) is a possible source that heats up the intergalactic medium (IGM) resulting in the rise in background temperature. Dark matter annihilation, decay and even the baryon-dark matter interaction could also induce the larger than the expected separation between $T_s$ and $T_{\gamma}$ \cite{rdi,legal_2,atri}. 
	%In the present work, we address the effects of possible decay of heavy dark matter or HDM on the 21-cm absorption temperature signal. In addition, the influence of PBH evaporation and the baryon-DM interactions are also included in a single framework to study the 21-cm absorption temperature and its evolution. 
	%In the present work, we investigate the combined effect of the HDM decay, PBH evaporation and the baryon-DM interaction in the framework of the global 21-cm signature.
	
	The hydrogen atom has two hyperfine electronic ground states namely a singlet spin 0 state and a triplet spin 1 state. The transition from the higher energy triplet state to the lower energy singlet produces the 21-cm radio spectral line which is described by the brightness temperature $T_{21}$. The expression of the brightness temperature at different redshift $z$ is given by,
	\begin{equation}
		T_{21}=\dfrac{T_s-T_{\gamma}}{1+z}\left(1-e^{-\tau(z)}\right)
		\label{eq:t21}
	\end{equation}
	where, $T_s$ is the spin temperature at redshift $z$ and $\tau(z)$ is the optical depth of the medium, given by \cite{munoz},
	\begin{equation}
		\tau(z) = \dfrac{3}{32 \pi}\dfrac{T_{\star}}{T_s}n_{\rm HI} \lambda_{21}^3\dfrac{A_{10}}{H(z)+(1+z)\delta_r v_r}.
		\label{eq:tau}
	\end{equation}
	In the above expression (Eq.~\ref{eq:tau}), $T_{\star}$ ($=hc/k_B \lambda_{21}=0.068$ K), $A_{10}=2.85\times 10^{-15}\,{\rm s^{-1}}$ is the Einstein coefficient \cite{yacine,hyrec11}, $\lambda_{21}\approx 21$ cm and $\delta_r v_r$ is the radial gradient of the peculiar velocity.
	
	The spin temperature ($T_s$) is characterized as, 
	\begin{equation}
		\frac{n_1}{n_0}=3 \exp{-\frac{T_{\star}}{T_s}},
	\end{equation}
	where $n_0$ and $n_1$ are the number densities of neutral hydrogen atoms at singlet spin-0 and triplet spin-1 states respectively. The resonant scattering of Lyman$\alpha$ (Ly$\alpha$) photons and the background photons modify the spin temperature. On the other hand, the combined effects of HDM decay, Hawking radiation and baryon-DM scattering also modify $T_s$ remarkably. The expression of $T_s$ as function of cosmic microwave background (CMB) temperature $T_{\gamma}$ ($T_{\gamma}=2.725 (1+z)$ K) and baryon temperature $T_b$ is given by,
	\begin{equation}
		T_s = \dfrac{T_{\gamma}+y_c T_b+y_{\rm Ly\alpha} T_{\rm Ly\alpha}}
		{1+y_c+y_{\rm Ly\alpha}},
		\label{eq:tspin}
	\end{equation}
	where, the term $y_{\rm Ly\alpha}$ arises due to the Wouthuysen-Field effect. In the above, $T_{\rm Ly\alpha}$ and $y_c$ are the Lyman-$\alpha$ background temperature and the collisional coupling parameter respectively \cite{BH_21cm_1}. The coefficients $y_c$ and $y_{\rm Ly\alpha}$ are given by $y_c=\frac{C_{10}T_{\star}}{A_{10} T_b}$ and $y_{\rm Ly\alpha}=\frac{P_{10}T_{\star}}{A_{10} T_{\rm Ly\alpha}}e^{0.3 \times (1+z)^{1/2} T_b^{-2/3} \left(1+\frac{0.4}{T_b}\right)^{-1}}$ \cite{BH_21cm_2,Yuan_2010,Kuhlen_2006}. Here, $P_{10}$ is the deexcitation rate due to Ly$\alpha$ given by $P_{10}\approx1.3\times 10^{-21}S_{\alpha}J_{-21}\,{\rm s^{-1}}$ where $C_{10}$ is the collision deexcitation rate, $S_{\alpha}$ is the spectral distortion factor \cite{salpha} and $J_{-21}$ represents the Lyman-$\alpha$ background intensity \cite{jalpha}.
	
	As mentioned earlier, energy injection is also possible in case of the decay of heavy primordial dark matter. Such dark matter can be gravitationally produced in the early Universe or it can also be produced during the reheating phase after inflation. Such a heavy dark matter can undergo cascading decay processes through hadronic or leptonic (or both) channels to produce ultra high energy (UHE) neutrinos (and other particles). The UHE neutrinos can be detected by Km$^2$ detectors such as IceCube and IceCube had reported neutrino signals in the PeV energy range. Such possibilities are earlier explored in Ref.~\cite{mpandey}.
	
	In this work, we study possible multimessenger signals of heavy dark matter whereby the two effects namely the neutrinos from heavy dark matter decay and the possible effects on 21-cm signal at reionization epoch due to the heat transfer following the decay process are addressed. In the present work, we consider two additional effects on the evolution of baryon temperature $T_b$, while computing the temperature of the 21-cm line at dark ages and reionization epoch. One is the evaporation of primordial black holes or PBHs and the other is the effects of the dark matter - baryon collision. 
		
	In the present calculations, the baryon-DM interaction cross-section ($\bar{\sigma}$) is parameterized as $\bar{\sigma}=\sigma_0 v^{-4}$ \cite{munoz}, where the lighter dark matter is considered to be model-independent \cite{IDM_1,IDM_2,Cheng:2002ej,servant_tait,Hooper:2007gi,Majumdar:2003dj,wimpfimp}. Velocity ($v$) dependent choice of this kind for dark matter-baryon scattering 
	cross-section has also been addressed earlier \cite{U66}. 
	Writing the cross-section $\sigma$ in a power law form as 
	$\sigma = \sigma_0 v^n$ where $v$ is given in the units of $c$, the velocity
	of light, the different values of the index $n$ are studied. For example,
	the case $n=-4$ is motivated by millicharged dark matter \cite{U67,U68}. Also $n = \pm 2$ appears relevant if dark matter has electric or 
	magnetic dipole moment and $n = \pm 1, 0$ are suitable for the scattering 
	involving Yukawa potential \cite{U69}. The proposition of the 
	form $\sigma = \sigma_0 v^{-4}$ in relation to the EDGES observation 
	are discussed in Ref.~\cite{rennan_3GeV} as also in Ref.~\cite{U66}.
	The cross-section $\sigma_0$ can be calculated by evaluating relevant 
	Feynman diagram of the scattering process for a given particle dark matter
	candidate. But here, we have considered a model independent WIMP like 
	cold dark matter of mass in the range of a few hundred MeV to few GeV,
	that scatters
	off baryons and thus affects the evolution of baryon temperature 
	during dark ages
	and reionization era and for this purpose we adopt $\sigma_0 = 10^{-41}$ 
	cm$^2$. This is in the ballpark of direct detection cross-section limits 
	of dark matter direct detection experiments. In the case of PBH heating, only the effect of the Hawking radiation is taken into account. %We address both the upper and lower bound of the initial mass fraction of PBHs ($\beta_{\rm BH}$) for a wide range of PBH mass and compare the upper bound with the same as obtained from the work of \citet{BH_21cm_3,BH_21cm_2}.
	It has already been mentioned that, in this work, we consider two types of dark matter. One is a dark matter in the mass range of hundreds of MeV to few GeV, that interacts with baryons and affects temperature $T_{21}$ of 21-cm line, while the other is a heavy dark matter that may have been produced after the inflationary epoch and can undergo cascading decay to produce leptons and injects energy into the system. The latter also can influence $T_{21}$, while the former accounts for most of the dark matter in the Universe and the fraction of heavy dark matter is negligibly small.
	
	The paper is organized as follows. Section~\ref{sec:HDM} describes the heavy dark matter decay and production of PeV neutrinos. Here IceCube neutrinos are assumed to be originated from heavy dark matter decay. In Section~\ref{sec:T_evol}, the formalism for the evolution of baryon temperature $T_b$ and dark matter temperature $T_{\chi}$ are described. Here both the effects of heavy dark matter decay and PBH evaporation are considered. The results and corresponding plots are described in Section~\ref{sec:result}. Finally, in Section~\ref{sec:conc}, the concluding remarks are given.
	
\section{\label{sec:HDM} Heavy Dark Matter Decay and Neutrino Production}
	The decay of heavy dark matter (HDM) having mass significantly higher than the electroweak scale takes place via the cascading of QCD partons \cite{kuz,bera,bera1,bera2}. We used only the hadronic decay channel ($\chi\rightarrow q\bar{q}$) in our calculation as the contribution of the leptonic decay channel ($\chi\rightarrow l\bar{l}$) is much smaller than the hadronic channel \cite{mpandey,kuz}. The total spectrum at scale $s=M_{\rm HDM}^2$ can be obtained by summing over all parton fragmentation function $D^{h} (x,s)$ from the decay of particle $h$. Here $x=\frac{2E}{M_{\rm HDM}}$ is the dimensionless energy. % and $i=$(quarks, gluons). 
	In our current analysis, only muon decay is considered in this calculation as the total contribution of the other mesons is negligible ($<10\%$) \cite{kuz,bera1}.
	
	The spectra of neutrinos ($\nu$), photons ($\gamma$) and electrons ($e$) are written as \cite{kuz},
	\begin{equation}
		\frac {dN_{\nu}}{dx}=2R \int_{xR}^{1} \frac{dy}{y} D^{\pi^{\pm}}(y)+2\int_{x}^{1}\frac{dz}{z}f_{\nu_i}\left(\frac{y}{z}\right)D^{\pi^{\pm}}(z),
		\label{form1}
	\end{equation}
	\begin{equation}
		\frac {dN_{\gamma}} {dx} = 2 \int_{x}^{1} \frac{dz}{z} D^{\pi^0}(z),
		\label{eq:gamma}
	\end{equation}
	\begin{equation}
		\frac {dN_{e}}{dx}=2R \int_{x}^{1} \frac{dy}{y}\left(\frac{5}{3}+3y^2+\frac{4}{3}y^3\right)\int_{x/y}^{x/(ry)}\frac{dz}{z}D^{\pi^{\pm}}(z).
		\label{eq:electron}
	\end{equation}
	In the above equations (Eqs.~\ref{form1}, \ref{eq:gamma} and \ref{eq:electron}), we can define $D^{\pi} (x,s)$ as $D^{\pi} \equiv [D_{q}^{\pi} (x,s) + 
	D_{g}^{\pi} (x,s)]$, where $x=\frac{2E}{M_{\rm HDM}}$ and $R = \displaystyle\frac {1} {1-r}$, where 
	$r = (m_\mu/m_\pi)^2 \approx 0.573$. The functions $f_{\nu_i} (x)$ are adopted from Ref.~\cite{kelner}
	\begin{eqnarray}
		f_{\nu_i} (x) &=& g_{\nu_i} (x) \Theta (x-r) +(h_{\nu_i}^{(1)} (x) + 
		h_{\nu_i}^{(2)} (x))\Theta(r-x) \,\, , \nonumber\\
		g_{\nu_\mu} (x) &=& \displaystyle\frac {3-2r} {9(1-r)^2} (9x^2 - 6\ln{x} -4x^3 -5), \nonumber\\
		h_{\nu_\mu}^{(1)} (x) &=& \displaystyle\frac {3-2r} {9(1-r)^2} (9r^2 - 6\ln{r} -
		4r^3 -5), \nonumber\\
		h_{\nu_\mu}^{(2)} (x) &=& \displaystyle\frac {(1+2r)(r-x)} {9r^2} [9(r+x) - 
		4(r^2+rx+x^2)], \nonumber\\
		g_{\nu_e} (x) &=& \displaystyle\frac {2} {3(1-r)^2} [(1-x) (6(1-x)^2 + 
		r(5 + 5x - 4x^2)) + 6r\ln{x}],\nonumber\\
		h_{\nu_e}^{(1)} (x) &=& \displaystyle\frac {2} {3(1-r)^2} [(1-r)
		(6-7r+11r^2-4r^3) + 6r \ln{r}], \nonumber\\
		h_{\nu_e}^{(2)} (x) &=& \displaystyle\frac {2(r-x)} {3r^2} (7r^2 - 4r^3 +7xr 
		-4xr^2 - 2x^2 - 4x^2r).
		\label{form2}
	\end{eqnarray}
	
	In this work, the neutrino spectra are obtained by computing Eqs.~\ref{form1} and \ref{form2} for different values of $M_{\rm HDM}$. The photon and electron spectra also can be compute using Eqs.~\ref{eq:gamma} -- \ref{form2}. The neutrino spectrum $\frac{dN_{\nu}}{dx}$ in Eq.~\ref{form1} represents the combination of neutrinos and antineutrinos spectrum of all three flavours ($\nu_e$, $\nu_{\mu}$ and $\nu_{\tau}$) where the neutrinos are considered to be produced from pion decay in the ratio 1:2:0 at the source.
	
	The extragalctic counterpart of	muon neutrino flux from heavy dark matter decay is given by \cite{kuz}, 
	\begin{equation}
		\frac{d\Phi_{\rm EG}}{dE} (E_\nu) = \frac{\mathcal{K}}{4\pi M_{\rm HDM}} \int_{0}^{\infty}\frac{\rho_0 c /H_0}{\sqrt{\Omega_m (1+z^3) + (1-\Omega_m)}} \frac{dN_{\nu_{\mu}}}{dE} [E(1+z)] dz.
		\label{form3}
	\end{equation}
	In the above equation (Eq.~\ref{form3}), $\mathcal{K}=f_{\rm HDM}\Gamma$, where $f_{\rm HDM}$ is the fraction of the heavy dark matter and $\Gamma$ is the decay width. The Hubble radius (the proper radius of the Hubble sphere) is given as $c/H_0$, where $c/H_0 = 1.37 \times 10^{28}$ cm. The average dark matter density of the Universe at $z=0$ (the present epoch) is denoted as $\rho_0 (= 1.15 \times 10^{-6}$ GeV/cm$^3$) and $\Omega_m = 0.316$ where $\Omega_m$ represents the contribution of the total dark matter density normalized to the critical density of the Universe. The neutrinos oscillate from one flavour to another during their propagation from the source to the Earth.
	But given the astronomical distance they traverse before reaching the Earth, the neutrinos originated with a flavour ratio $\nu_e:\nu_{\mu}:\nu_{\tau}=1:2:0$ and reaches the Earth (as the oscillation part is averaged out) with a flavour ratio $1:1:1$. Thus the muon neutrino spectrum $\frac{dN_{\nu_{\mu}}}{dE}=\frac{1}{3}\frac{dN_{\nu}}{dE}$ on reaching the Earth.
	
	In addition to the extragalactic part, the galactic $\nu_{\mu}$ flux from similar decay is given by \cite{kuz}
	\begin{equation}
		\frac{d\Phi_{\rm G}}{dE} (E_\nu) = \frac{\mathcal{K}}{4\pi M_{\rm HDM}} \int_{V} \frac{\rho_\chi (R[r])}{4\pi r^2} \frac{dN_{\nu_{\mu}}}{dE}(E,l,b) dV,
		\label{form4}
	\end{equation}
	where $l$ and $b$ denote the Galactic coordinates. The density of dark matter at a distance $R$ from the Galactic centre (GC) is denoted as $\rho_{\chi} (R[r])$ and $r$ is the distance from the Earth. In our calculation, we have taken into account the Navarro-Frenk-White (NFW) profile for the dark matter density and the integration is performed over the Milky Way halo for which $R_{\rm max}$ is considered to be 260 Kpc. 
	
	The total flux, which is also referred to as the total theoretical flux ($\phi^{\rm Th}$), at energy $E_{\nu}$ can be obtained by considering the contribution of both the galactic and the extragalactic $\nu_{\mu}$ flux as
	\begin{equation}
		\phi^{\rm Th} (E_\nu) = \frac{d\Phi_{\rm EG}}{dE} (E_{\nu}) + \frac{d\Phi_{\rm G}}{dE} (E_\nu).
		\label{form5}
	\end{equation}
	The quantities $\mathcal{K}$ and $M_{\rm HDM}$ are two parameters to be obtained by comparing and fitting the IceCube experimental data with $\phi^{\rm Th} (E_\nu)$.

\section{\label{sec:T_evol} Heavy Dark Matter Decay and Baryon Temperature Evolution in Dark Ages}
	The energy injection rate per unit volume due to decay of HDM is given by
	\begin{equation}
		\left.\dfrac{{\rm d} E}{{\rm d}V {\rm d}t}\right|_{\rm{HDM}}=\rho_{\chi} f_{\rm HDM} \Gamma,
		\label{eq:hdm_inj_0}
	\end{equation}
	where, as mentioned in Section~\ref{sec:HDM}, $\Gamma$ is the decay width, $f_{\rm HDM}$ is the fraction of heavy dark matter and $\rho_{\chi}$ is the total dark matter density of the Universe and $\rho_{\chi}=\rho_{\chi,0} (1+z)^3$ where $\rho_{\chi,0}$ is the total dark matter density in the current epoch. With $\mathcal{K}=f_{\rm HDM} \Gamma$, the above equation is reduced to  
	\begin{equation}
		\left.\dfrac{{\rm d} E}{{\rm d}V {\rm d}t}\right|_{\rm{HDM}}=\rho_{\chi}\mathcal{K}.
		\label{eq:hdm_inj}
	\end{equation}

	In our analysis, the thermal evolution of the charge-neutral Universe is studied by evolving the dark matter temperature ($T_{\chi}$) and the baryon temperature ($T_b$) with cosmological redshift $z$. After incorporating the effects of energy injection due to HDM decay as also due to PBH evaporation and the baryon-DM interaction, the evolution equations ($T_{\chi}$ and $T_{b}$ of Ref.~\cite{munoz}) take the form \cite{corr_equs,munoz,BH_21cm_1,BH_21cm_2,amar21},
	\begin{equation}
		(1+z)\frac{{\rm d} T_\chi}{{\rm d} z} = 2 T_\chi - \frac{2 \dot{Q}_\chi}{3 H(z)}, 
		\label{eq:T_chi}
	\end{equation}
	\begin{equation}
		(1+z)\frac{{\rm d} T_b}{{\rm d} z} = 2 T_b + \frac{\Gamma_c}{H(z)}
		(T_b - T_{\gamma})-\frac{2 \dot{Q}_b}{ 3 H(z)}-\frac{2}{3 k_B H(z)} \frac{K_{\rm BH}+K_{\rm HDM}}{1+f_{\rm He}+x_e}, 
		\label{eq:T_b}
	\end{equation}
	 where $\Gamma_c$ is the rate of Compton interaction, given by $\Gamma_c=\frac{8\sigma_T a_r T^4_{\gamma}x_e}{3(1+f_{\rm He}+x_e)m_e c}$. $\sigma_r$ and $a_T$ are the radiation constant and the Thomson scattering cross-section respectively and $x_e$ and $f_{\rm He}$ are the ionization fraction $x_e$ ($=n_e/n_H$, where $n_e$ and $n_H$ are the number density of free electron and hydrogen respectively) and the He fraction respectively. The terms $\dot{Q}_b$ and $\dot{Q}_{\chi}$ denote the heating rate of baryon and dark matter fluid respectively due to baryon-dark matter interaction. In the present calculation  $\dot{Q}_b$ and $\dot{Q}_{\chi}$ are calculated as described in Ref.~\cite{munoz}. The last term of the Eq.~\ref{eq:T_b} represents additional contributions to the evolution of baryon temperature $T_b$ due to heavy dark matter (HDM) decay \cite{Mitridate:2018iag,BH_21cm_1} and PBH evaporation \cite{BH_21cm_1,BH_21cm_2,amar21}. The quantities $K_{\rm BH}$ and $K_{\rm HDM}$ are defined as
	 \begin{equation}
	 	K_{\rm HDM(BH)}=\chi_h f_{\rm HDM(BH)}(z) \frac{1}{n_b} \times \left.\dfrac{{\rm d} E}{{\rm d}V {\rm d}t}\right|_{\rm{HDM(BH)}}. \label{KBH}
	 \end{equation}
 	Here $\chi_h=(1+2 x_e)/3$ is the fraction of energy deposited in the system due to heating \cite{BH_21cm_2,chen,BH_21cm_4,PhysRevD.76.061301,Furlanetto:2006wp}, $f_{\rm HDM(BH)}(z)$ is the ratio of the total amount of energy deposited and the injected energy due to HDM decay (PBH evaporation) \cite{corr_equs,fcz001,fcz002,fcz003,fcz004}, $n_b$ is the baryon number density.
	
	The ionization fraction $x_e$ of the Universe depends significantly on the amount of energy deposited in the medium. The variation of the ionization fraction $x_e$ with redshift $z$  given by \cite{munoz, BH_21cm_2, amar21},	
	\begin{equation}
		\frac{{\rm d} x_e}{{\rm d} z} = \frac{1}{(1+z)\,H(z)}\left[I_{\rm Re}(z)-
		I_{\rm Ion}(z)-(I_{\rm BH}(z)+I_{\rm HDM}(z))\right], 
		\label{eq:xe}
	\end{equation}
	where, $I_{\rm Re}(z)$ is the standard recombination and $I_{\rm Ion}(z)$ is the ionization rate respectively \cite{yacine,hyrec11,munoz,amar21,amar21_1}. As both terms ($I_{\rm Re}(z)$ and $I_{\rm Ion}(z)$) are functions of $T_b$ and $T_{\gamma}$ \cite{peeble,hyrec11,hummer,pequignot,seager,pequignot,BH_21cm_5,amar21,amar21_1}, the term $x_e$ depends on baryon temperature and DM temperature simultaneously. The quantities $I_{\rm BH}$ and $I_{\rm HDM}$ appearing in the evolution equation of the ionization fraction (Eq.~\ref{eq:xe}) are described as
	\begin{equation}
		 I_{\rm HDM(BH)}=\chi_i f_{\rm HDM(BH)}(z) \frac{1}{n_b} \frac{1}{E_0}\times \left.
		 \dfrac{{\rm d} E}{{\rm d}V {\rm d}t}\right|_{\rm HDM(BH)}. \label{IBH}
	\end{equation}
	where $E_0$ is the amplitude of the ground state energy of hydrogen atom, $\chi_i=(1-x_e)/3$ is the fractions of the energy deposited in the form of ionization \cite{BH_21cm_2,chen,BH_21cm_4,PhysRevD.76.061301,Furlanetto:2006wp}. Note that, in the case of HDM decay, the term $f_{\rm HDM}(z)$ is estimated by incorporating only the photon and electron spectra as obtained by the Eqs.~\ref{eq:gamma} and \ref{eq:electron} (see section~\ref{sec:HDM}) ~\cite{corr_equs,fcz001,fcz002,fcz003,fcz004} (the energy deposition into the intergalactic medium by neutrinos is negligible \cite{BH_21cm_1,fcz001,fcz002,fcz003,fcz004}). 	
	At higher redshift and energy, almost entire flux of electrons transfer to photons by the process of inverse Compton scatter (ICS) and above the threshold of pair-production, the photons efficiently produce $e^{-}e^{+}$ pair. Those electrons and positrons will rapidly reduce to low energy $e^{-}$ and $e^{+}$ via ICS and cascade. As a consequence, in the calculation of $f_{\rm HDM}(z)$, the transfer function does not change significantly with increasing DM mass at higher values. Moreover, it can be noticed that, the available grid data of energy transfer function \cite{fcz001,fcz002} is almost constant for $\gtrapprox 10^3$ GeV for fixed values of input and output redshift \cite{corr_equs,fcz001,fcz002,fcz003,fcz004}. So, in the case of HDM with different masses, we use the transfer function for $10^{12.75}$ eV in our calculations (the electron and photon spectrum are estimated from Eqs.~\ref{eq:gamma} and \ref{eq:electron} of Section~\ref{sec:HDM}).
	
	The baryon-DM fluid interaction depends on the drag term $V_{\chi b}$ \cite{munoz}. The evolution of $V_{\chi b}$ is described as \cite{munoz},
	\begin{equation}
		\frac{{\rm d} V_{\chi b}}{{\rm d} z} = \frac{V_{\chi b}}{1+z}+
		\frac{1}{(1+z) H(z)}\dfrac{\rho_m \sigma_0}{m_b + m_{\chi}} \dfrac{1}{V^2_{\chi b}} F(r), \label{eq:V_chib}
	\end{equation}
	where, $F(r)={\rm erf}\left(r/\sqrt{2}\right)-\sqrt{2/\pi}r e^{-r^2/2}$ ($\rm erf$ represents the error function), $r=V_{\chi b}/u_{\rm th}$ and $\sigma_{41}=\frac{\sigma_0}{10^{-41} {\rm cm^2}}$. The term $u_{\rm th}^2$ is the variance of the relative thermal motion.
	
	It is to be noted that two categories of dark matter are considered in this work. One is the heavy dark matter produced in the early Universe, the decay of which inject energy into the system and produce UHE neutrinos, while the other is the usual WIMP dark matter that scatters with baryons and exchange heat. Both the processes influence the temperature of the 21-cm line.
	
	In the case of PBH evaporation, the radiation contains several particle species \citep{PhysRevD.41.3052,PhysRevD.94.044029,BH_21cm_2,BH_F}. The rate of the energy injection due to Hawking radiation is written as \cite{BH_21cm_2,amar21},
	\begin{equation}
		\left.\dfrac{{\rm d} E}{{\rm d}V {\rm d}t}\right|_{\rm{BH}}=-\dfrac{{\rm d} M_{\rm{BH}}}{{\rm d} t} n_{\rm BH}(z)
		\label{dedvdt}
	\end{equation}
	where, $n_{\rm{BH}}(z)$ is the PBH number density at redshift $z$, and is given by \cite{BH_21cm_2},
	\begin{eqnarray}
		n_{\rm{BH}}(z)&=&\beta_{\rm BH}\left(\dfrac{1+z}{1+z_{\rm eq}}\right)^3 \dfrac{\rho_{\rm c,eq}}{\mathcal{M}_{{\rm BH}}} \left(\dfrac{\mathcal{M}_{\rm H,eq}}{\mathcal{M}_{\rm H}}\right)^{1/2} \left(\dfrac{g^i_{\star}}{g^{\rm eq}_{\star}}\right)^{1/12}\nonumber\\
		&\approx&1.46 \times 10^{-4}\beta_{\rm BH} \left(1+z\right)^3 \left(\dfrac{\mathcal{M}_{{\rm BH}}}{\rm g}\right)^{-3/2} {\rm cm^{-3}}
		\label{nbh}
	\end{eqnarray}
	In the above, $\mathcal{M_{\rm BH}}$ is the mass of the PBH at the time of formation and $\mathcal{M_{\rm H}}$ is the horizon mass \cite{BH_21cm_2,BH_21cm_3,betabh}. The quantity $\beta_{\rm BH}$ represents the initial mass fraction of PBHs.
	
	A black hole of mass $M_{\rm{BH}}$ evaporates at the rate \cite{hawking,BH_21cm_1}
	\begin{equation}
		\dfrac{{\rm d}M_{\rm{BH}}}{{\rm d}t} \approx -5.34\times10^{25} \left(\sum_{i} \mathcal{F}_i\right) M_{\rm{BH}}^{-2} \,\,\rm{g/sec}
		\label{eq:PBH}
	\end{equation}
	where the coefficient $\mathcal{F}_i$ represents the evaporation fraction in the form of $i^{\rm th}$ particle. The energy injection due to the HDM decay is already shown in Eq.~\ref{eq:hdm_inj}.
	%The energy injection due to the HDM decay is given by,
	%\begin{equation}
	%	\left.\dfrac{{\rm d} E}{{\rm d}V {\rm d}t}\right|_{\rm{HDM}}=\rho_{\chi} f_{\rm HDM} \Gamma,
	%	\label{eq:hdm_inj}
	%\end{equation}
	%where $f_{\rm HDM}$ is the fraction of dark matter in the form of heavy dark matter and $\Gamma$ is the width of HDM decay. In the present work, we introduce a new parameter $\mathcal{K}=f_{\rm HDM} \Gamma_{\rm HDM}$. With this, Eq.~\ref{eq:hdm_inj} takes the form, $\left.\dfrac{{\rm d} E}{{\rm d}V {\rm d}t}\right|_{\rm{HDM}}=\rho_{\chi} \mathcal{K}$.
	
\section{\label{sec:result} Calculations and Results}
	\begin{figure}
		\centering{}
		\includegraphics[trim=0 20 0 60, clip, width=0.7\linewidth]{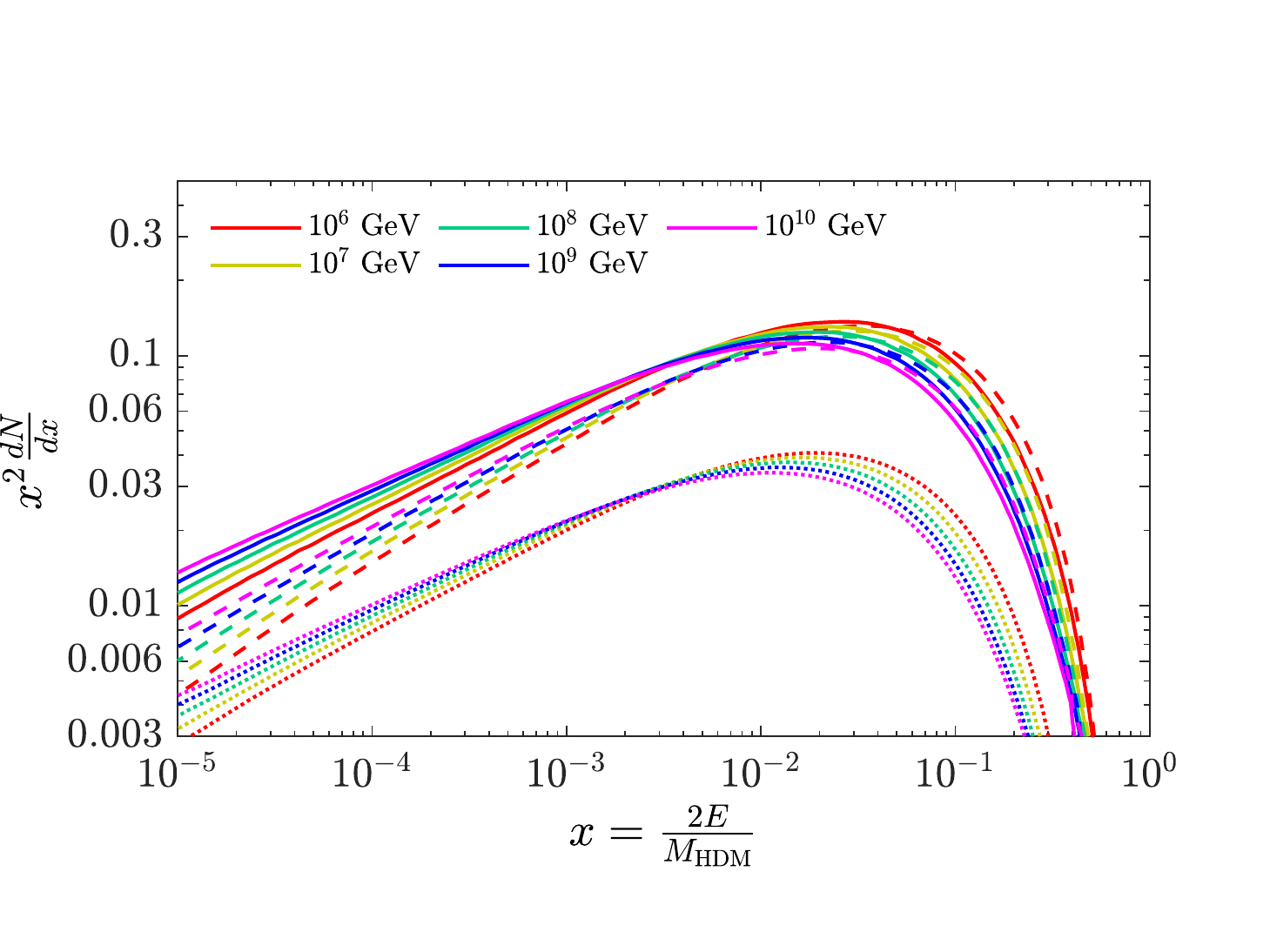}
		\caption{\label{fig:spectrum} Spectra of heavy dark matter decay for different dark matter masses. The solid lines represent the total neutrino spectra ($\nu_e+\nu_{\mu}+\nu_{\tau}$, $\nu+\bar{\nu}$) for different $M_{\rm HDM}$. The corresponding $\gamma$ and $e$ spectra are described by dashed and dotted lines respectively.}
	\end{figure}

	As mentioned earlier, in the present work, we explore the possible multimessenger effects due to heavy dark matter decay. One is the UHE $\nu$-signal at the IceCube detector and the other is the effect on 21-cm hydrogen line signal during dark ages leading to reionization epoch due to the energy injection following the decay processes.
	
	The computation of the neutrino spectrum is performed by numerically evaluating Eqs.~\ref{form1}--\ref{form2}. The calculations involve Monte Carlo simulation of numerical evolution of DGLAP equations for obtaining QCD spectrum. These computed neutrino spectra are then used to obtain neutrino flux using Eqs.~\ref{form3}--\ref{form5}. These are furnished in Fig.\ref{fig:spectrum}. In Fig.~\ref{fig:spectrum}, the total neutrino flux obtained from the above computations is shown for five chosen values of $M_{\rm HDM}$, the mass of heavy dark matter. These are shown in Fig.~\ref{fig:spectrum} using solid lines of different colours for different chosen $M_{\rm HDM}$ values. Spectra for $\gamma$ and $e^-$ (from HDM decay) are also shown in Fig.~\ref{fig:spectrum} for reference using coloured dashed lines.
	
	The experimental data are chosen from the 7.5 yr data corresponding to muon through going events. These events beyond the energy $\sim$120 TeV appear to describe a power law of High Energy Starting Events or HESE that include both track and shower events and are assumed to be obtained from diffuse extragalactic UHE neutrino flux \cite{Kopper:2017Df}. 
	
	\begin{figure}
		\centering{}
		\includegraphics[width=0.7\linewidth]{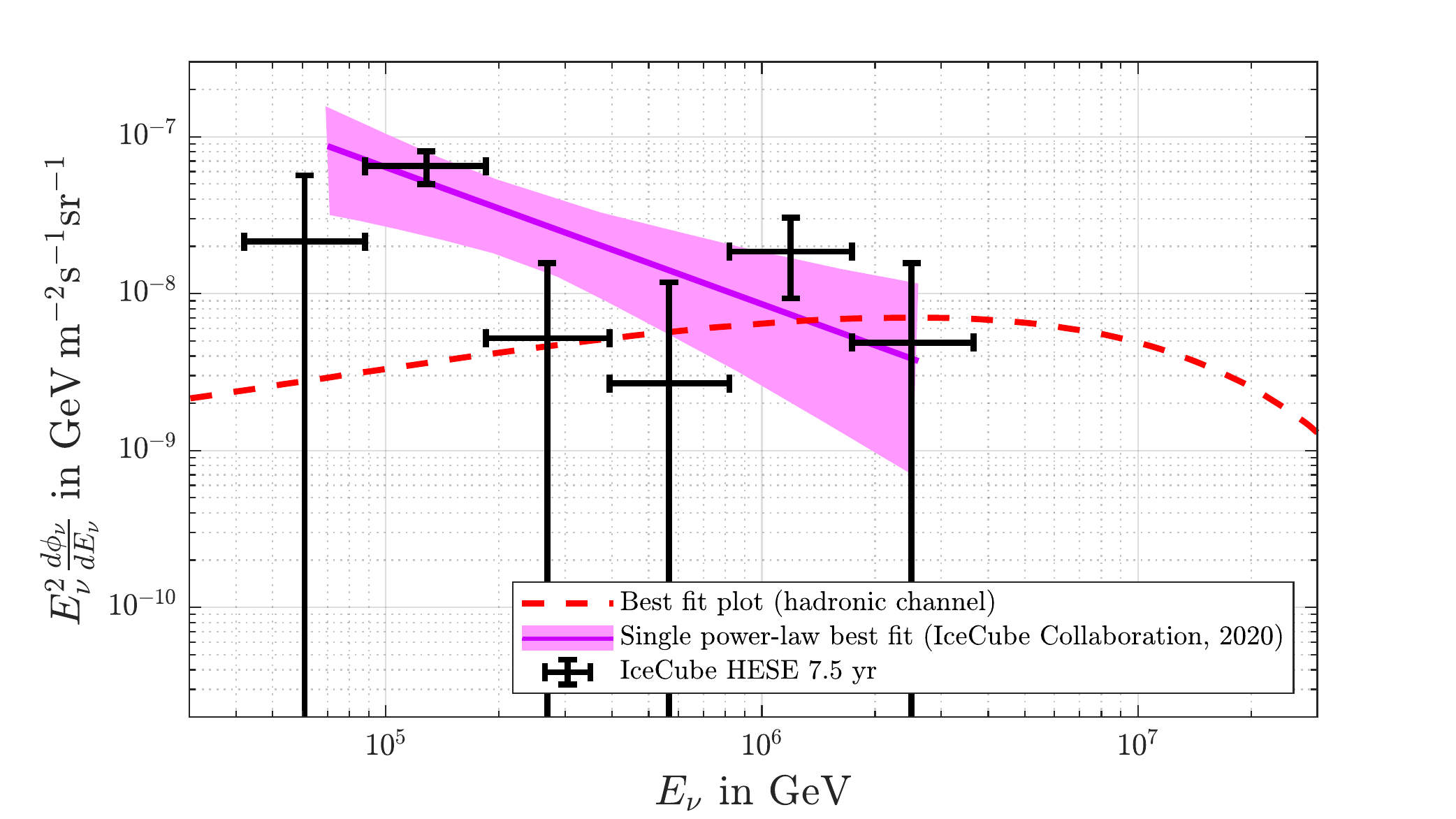}
		\caption{\label{fig:flx} Best fit $\nu_{\mu}$ spectrum from heavy dark matter decay.}
	\end{figure}
	In Fig.~\ref{fig:flx}, the through going best fit power law (different from HESE single power-law fit of $E^{-\gamma}$ with $\gamma \sim 2.9$) is shown along with its statistical fluctuation is shown by the pink band. The actual observed data points for through-going muon events (from 7.5 yr data) are shown in Fig.~\ref{fig:flx} with the error bars. We have chosen four data points in the neutrino energy region $2\times 10^5$ GeV -- $4 \times 10^6$ GeV. The other two data points in the lower energy regime in Fig.~\ref{fig:flx} belong to Astrophysical Neutrino events \cite{Chianese_2016} and hence are not considered in the present analysis. These four experimental data points are referred as $\phi^{\rm Ex}$.
	
\begin{figure*}
		\centering{}
		\begin{tabular}{cc}
			\includegraphics[width=0.48\textwidth]{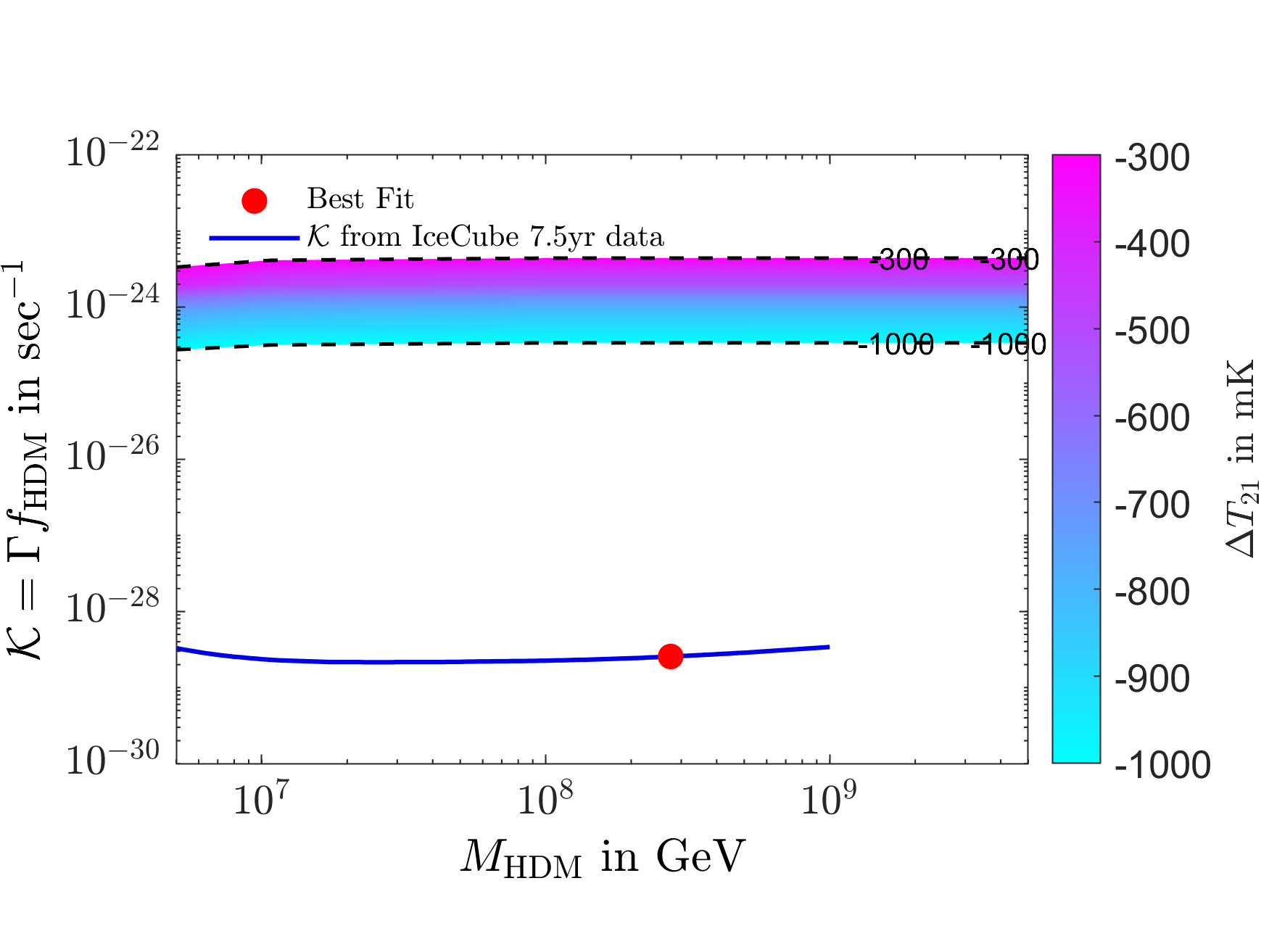}&
			\includegraphics[width=0.48\textwidth]{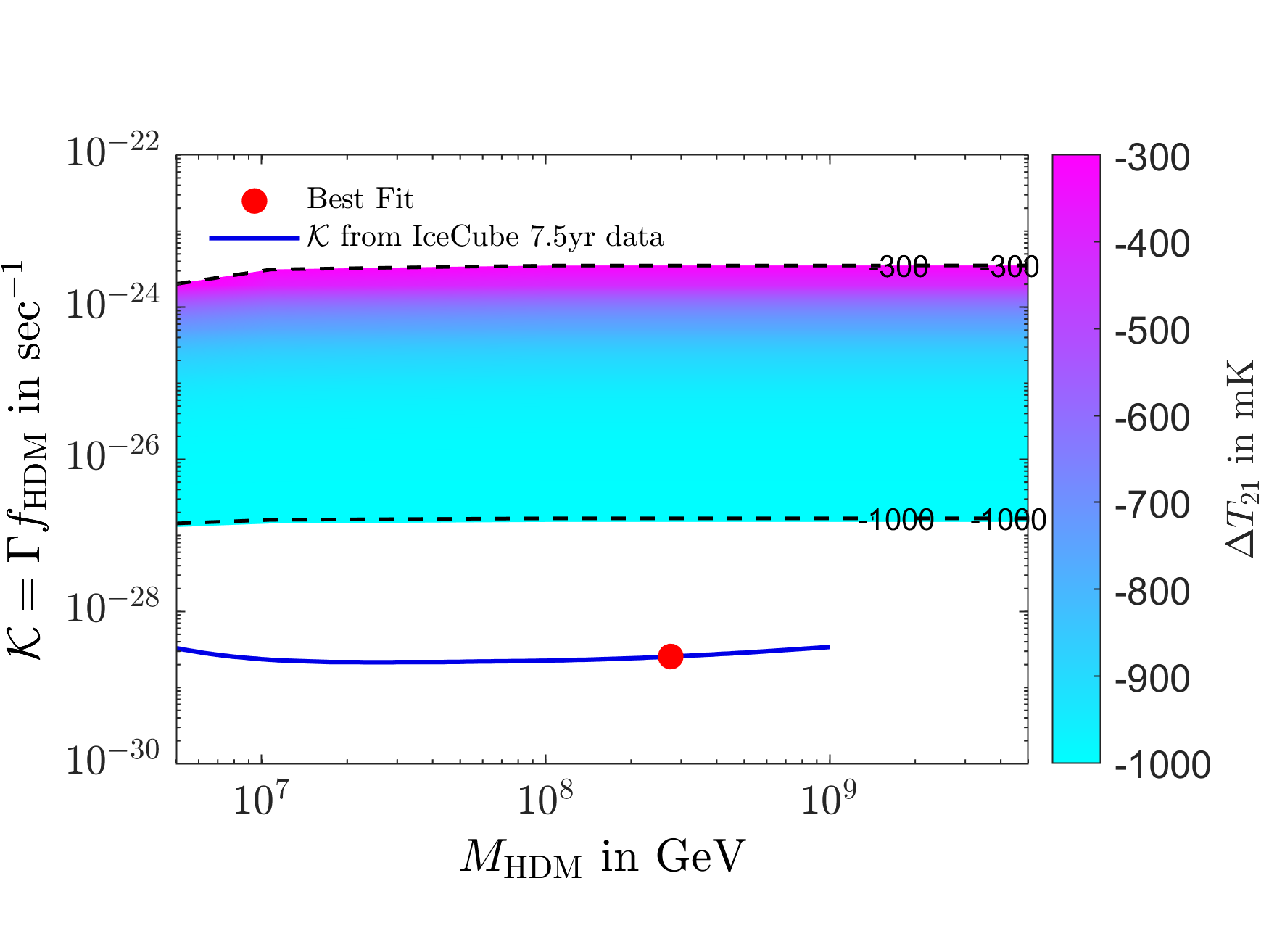}\\
			(a)&(b)\\
			\includegraphics[width=0.48\textwidth]{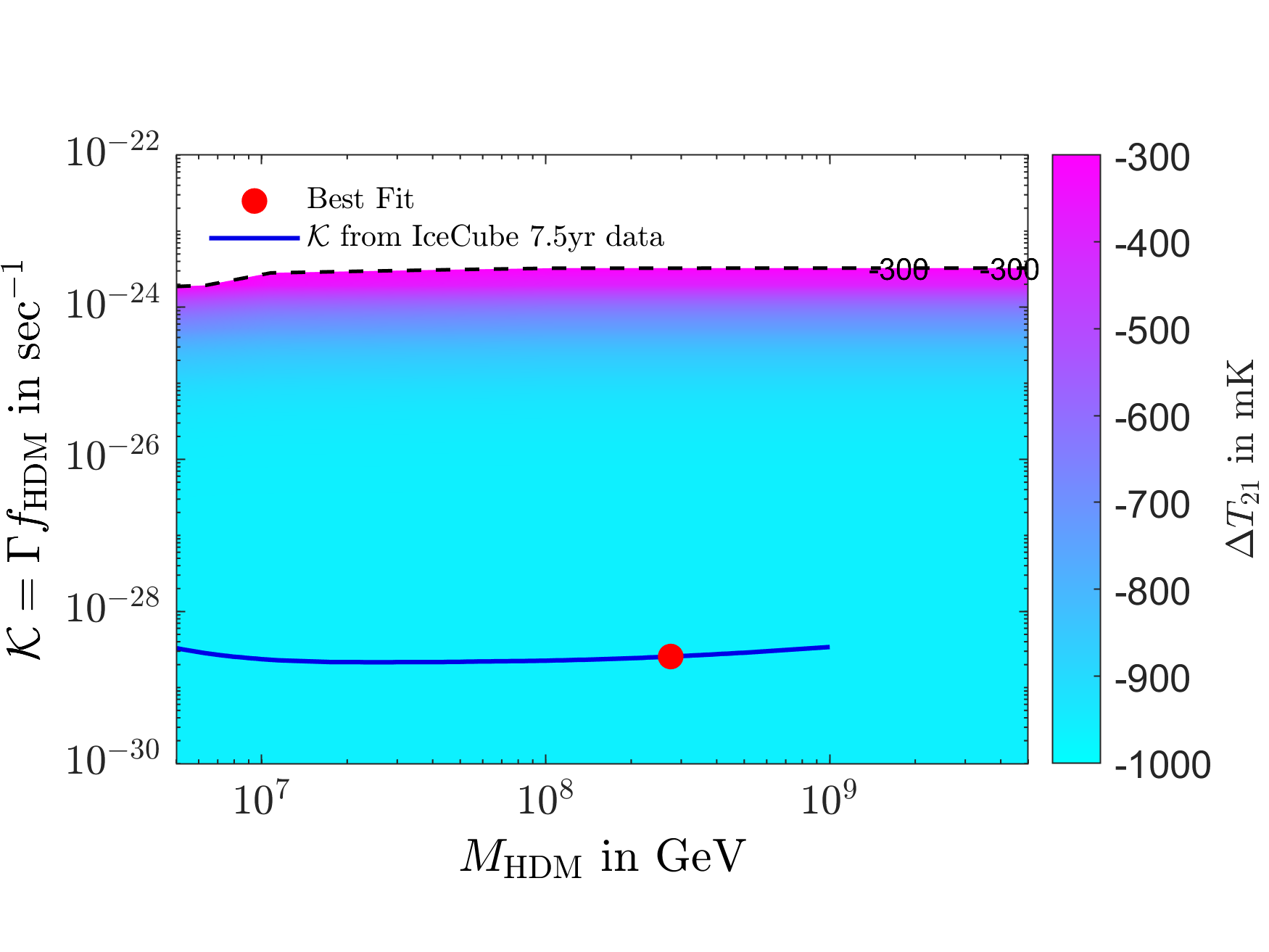}&
			\includegraphics[width=0.48\textwidth]{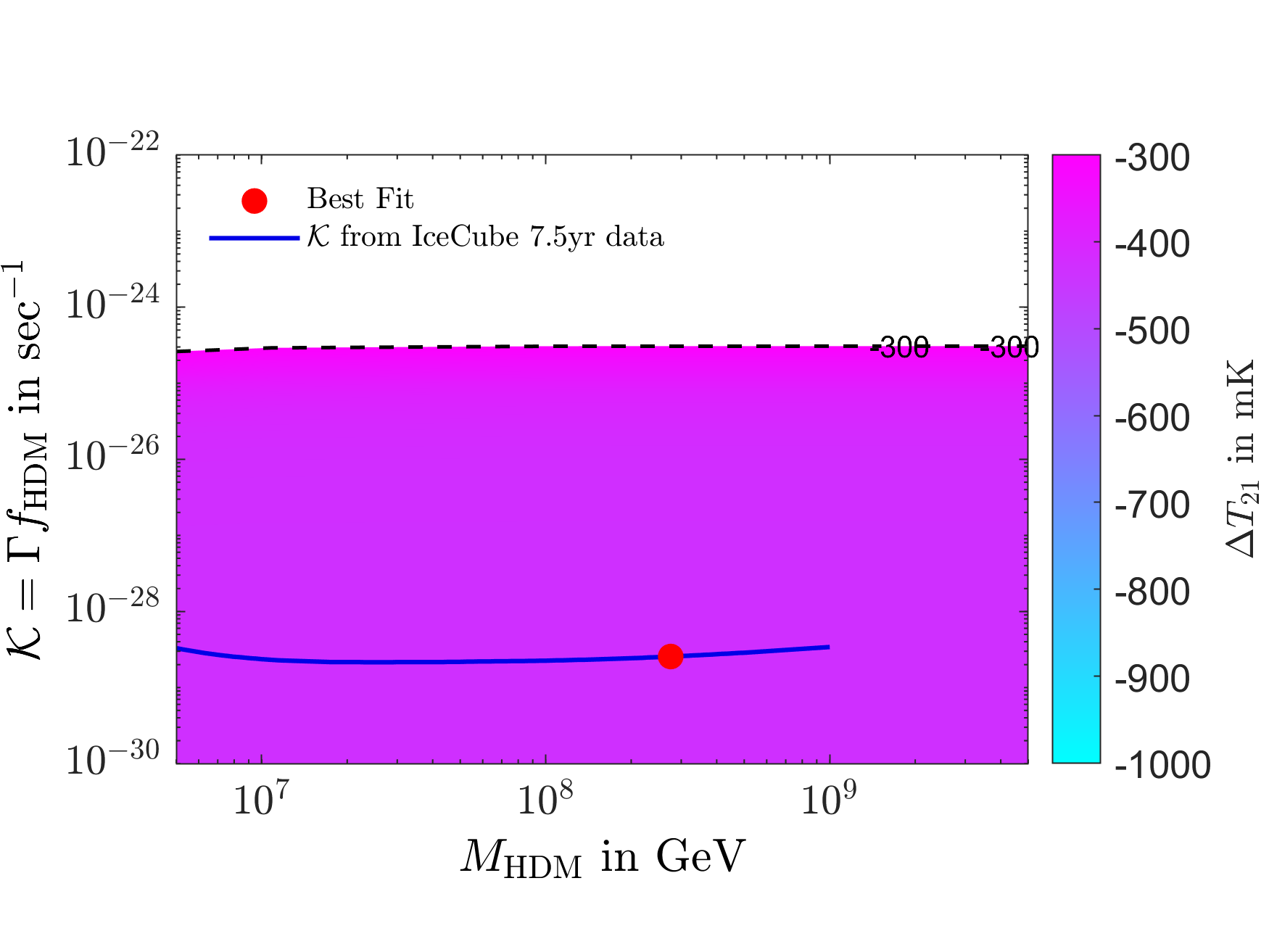}\\
			(c)&(d)\\
		\end{tabular}
		\caption{\label{fig:Mx_mathK} The allowed zone in the $\mathcal{K}$-$M_{\rm HDM}$ plane for (a) $m_{\chi}=0.3$ GeV, (b) $m_{\chi}=0.38$ GeV, (d) $m_{\chi}=0.4$ GeV and (d) $m_{\chi}=1.0$ GeV, where other variables are kept fixed at $\sigma_{41}=1$, $\mathcal{M_{\rm BH}}=10^{14}$g and $\beta_{\rm BH}=10^{-29}$. The upper and lower bounds are correspond to the $\Delta T_{21}=-300$ mK and $-1000$ mK respectively. The solid blue line indicates best fit values of $\mathcal{K}$ for different values of $M_{\rm HDM}$ (from IceCube 7.5 year up-going event data), while the red dot symbol specifies the best fit $\mathcal{K}$ for best fit $M_{\rm HDM}$.}
	\end{figure*}

	The parameters $\mathcal{K}$ and $M_{\rm HDM}$ are obtained by making a $\chi^2$ fit ($\chi^2$ minimization) of data $\phi^{\rm Ex}$ with $\phi^{\rm Th}$. The $\chi^2$ is defined as,
	\begin{equation}
		\chi^2 = \dfrac{1}{n}\sum_{i=1}^{n} \left(\frac{E_i^2 \phi_i^{\rm Th}-E_i^2 \phi_i^{\rm Ex}} {(\rm err)_i} \right)^2,
		\label{cal1}
	\end{equation}
	where the number of chosen points are $n(=4)$ and $E_i$ represents the energy corresponding to data point $i$. In Eq.~\ref{cal1}, $({\rm err})_i$ is the error of the $i^{\rm th}$ chosen experimental point. For this two-parameter $\chi^2$ fit, we consider only the hadronic decay channel as the contribution due to the leptonic decay channel for the chosen four data points is negligible \cite{mpandey}.
	From the analysis, the obtained best fit values of HDM mass ($M_{\rm HDM}$) and $\mathcal{K}$ are $M_{\rm HDM}=2.75\times 10^8$ GeV and $\mathcal{K}=2.56\times 10^{-29}\,{\rm sec^{-1}}$ respectively. The red dashed line in Fig.~\ref{fig:flx} indicates the calculated $\nu_{\mu}$ flux for the best fit values of $M_{\rm HDM}$ and $\mathcal{K}$. Moreover, we also estimate best fit values of $\mathcal{K}$ for different HDM masses (see Fig.~\ref{fig:Mx_mathK}).
	
	The other multimessenger effect considered here is related to exploring the influence of heavy dark matter decay on the global 21-cm signature, where the effects of PBH evaporation and the baryon-DM interaction have also been included. Two categories of dark matter are considered here. One is the possible decay of heavy dark matter to produce ultra-high energy neutrinos and their detection by IceCube, while the other is a WIMP type cold dark matter (CDM) interaction strength in the weak interaction regime. We assume that the fraction of DM in the form of HDM is very small in comparison to that of the lighter CDM-type dark matter. %Our aim in the current analysis is to investigate the allowed values $\mathcal{K}$ for heavy dark matter decay and its variation with other system parameters (i.e. $\mathcal{M}_{\rm BH}$, $m_{\chi}$, $\sigma_{41}$). 
	From Eqs.~\ref{eq:T_b}, \ref{KBH}, \ref{eq:hdm_inj}, it can be seen that the decay of heavy dark matter (and the PBH evaporation) contributes to the evolution of baryon temperature $T_b$ and consequently the 21-cm absorption temperature $T_{21}$. Also, the heavy dark matter cascading decay can produce UHE neutrinos that could be detected by km$^2$ detector such as IceCube. Here in this work, the best fit values for heavy dark matter decay width (in fact $\mathcal{K}=\Gamma f_{\rm HDM}$) is further constrained for different possible HDM masses $M_{\rm HDM}$ in case of different $m_{\chi}$ values by the EDGES result for 21-cm absorption line. While computing $T_{21}$, in addition to the effects of heavy dark matter decay, the collisional effects of baryons with CDM (that is assumed to account for almost all the dark matter in the Universe), the evaporation of primary black holes etc. By performing this analysis we also attempted to explore the contribution of a very small fraction of heavy dark matter through its decay and the contribution of the overwhelming CDM type lighter dark matter through their collisional effects with baryon while also including PBH evaporation contributions to 21-cm Hydrogen absorption line during reionization era. As mentioned, various constraints are estimated using the experimental results of the EDGES experiment ($T_{21} = -500^{+200}_{-500}$ at $z=17.2$), the brightness temperature at redshift $z=17.2$ is an important quantity in this analysis, represented by $\Delta T_{21}$.
	
	It is to be mentioned that, in the present analysis the evolution of temperature and other quantities are computed considering the redshift $z=1100$ as the initial epoch (similar to what described in Ref.~\cite{munoz}). Therefore, all the initial condition in the present calculations corresponds to the epoch of redshift $z=1100$. At that redshift ($z=1100$), the baryons are assumed to be tightly coupled with the background photons while, the temperature of the dark matter fluid is considered to be $T_{\chi}=0$ and the initial relative velocity $V_{\chi b}\sim 29$ km/s at that redshift ($z=1100$) for each combination of the parameters $m_{\chi}$, $\sigma_{41}$, $\mathcal{K}$, $M_{\rm HDM}$, $\mathcal{M_{\rm BH}}$ and $\beta_{\rm BH}$. It is to be mentioned that, even if a slightly warm dark matter candidate is taken into the account, the evolution remains almost the same (it is also tested in the work of Ref.~\cite{munoz}). Since we evolve the temperatures of two matter fluids (dark matter fluid and baryon fluid) simultaneously, the cooler fluid (dark matter fluid) tends to heat up at the expense of the temperature of the comparatively warm fluid (baryon fluid) as an outcome of the temperature exchange between them and the hotter fluid (baryonic fluid) cools down. Although this heating rate is essentially proportional to ($T_b-T_{\chi}$), the heating rate gets perturbed in presence of the drag term ($V_{\chi b}$). A detailed analysis regarding this thermal effect can be seen in Ref.~\cite{munoz}. On the other hand, the effects of PBH evaporation and HDM decay essentially raise the baryon temperature.
		
	We simultaneously solve numerically Eqs.~\ref{eq:T_chi} -- \ref{eq:hdm_inj} and obtain the baryon temperature. The spin temperature is then computed using Eq.~\ref{eq:tspin} where the effect of Ly$\alpha$ forest are also included. Finally $T_{21}$ is calculated using Eqs.~\ref{eq:t21}, \ref{eq:tau}. In Fig.~\ref{fig:Mx_mathK}, the allowed regions in the $\mathcal{K}$-$M_{\rm HDM}$ parameter plane are shown for different chosen values of DM mass $m_{\chi}$. The four plots in Fig.~\ref{fig:Mx_mathK} corresponds to (a) $m_{\chi}=0.3$ GeV, (b) $m_{\chi}=0.38$ GeV, (c) $m_{\chi}=0.4$ GeV and (d) $m_{\chi}=1.0$ GeV. The allowed region is estimated using the experimental excess of EDGES. The uppermost line of the shaded region (upper dashed black line) corresponds to $\Delta T_{21}=-300$ mK while the lower black dashed line corresponds to $\Delta T_{21}=-1000$ mK. The values of $\Delta T_{21}$ at any points between the upper and lower limits are indicated by the corresponding colour codes (see colour bar). The best fit values of $\mathcal{K}$ and $M_{\rm HDM}$ as obtained from the $\chi^2$ minimization fit of IceCube 7.5 yr through going data are shown in all the four plots of Fig.~\ref{fig:Mx_mathK}. The best fit values of $\mathcal{K}$ for every chosen values of $M_{\rm HDM}$ in the region $\sim 10^6$--$\sim 10^9$ GeV (as obtained from $\chi^2$ minimization of IceCube 7.5 yr through going data) are shown as a solid line in all the plots of Fig.~\ref{fig:Mx_mathK}.
	In the case of $m_{\chi}=0.3$ GeV, the allowed zone in $\mathcal{K}$-$M_{\rm HDM}$ plane lies far above the $\mathcal{K}$-best fit line. However, comparing the plots of Fig.~\ref{fig:Mx_mathK}(a, b, c, d), we see that, as $m_{\chi}$ increases, the lower bound of $\mathcal{K}$-$M_{\rm HDM}$ region falls rapidly, while the upper bound of the allowed region decreases gradually. It can be seen from Fig.~\ref{fig:Mx_mathK} that for $m_{\chi} \gtrapprox 0.4$ GeV, the best fit line lies within the $\mathcal{K}$-$M_{\rm HDM}$ allowed region. For all the cases, the values of PBH parameters are chosen to be $\mathcal{M_{\rm BH}}=10^{14}$g and $\beta_{\rm BH}=10^{-29}$, while $\sigma_{41}$ is fixed at 1. %We have also repeated the same analysis for $\mathcal{M_{\rm BH}}=10^{15}$g.% (graphs are not included in the manuscript). In that case, we found that the allowed region of $\mathcal{K}$ matches the best fit values at comparatively higher values of $m_{\chi}$.
		
	\begin{figure}
		\centering{}
		\includegraphics[width=0.7\linewidth]{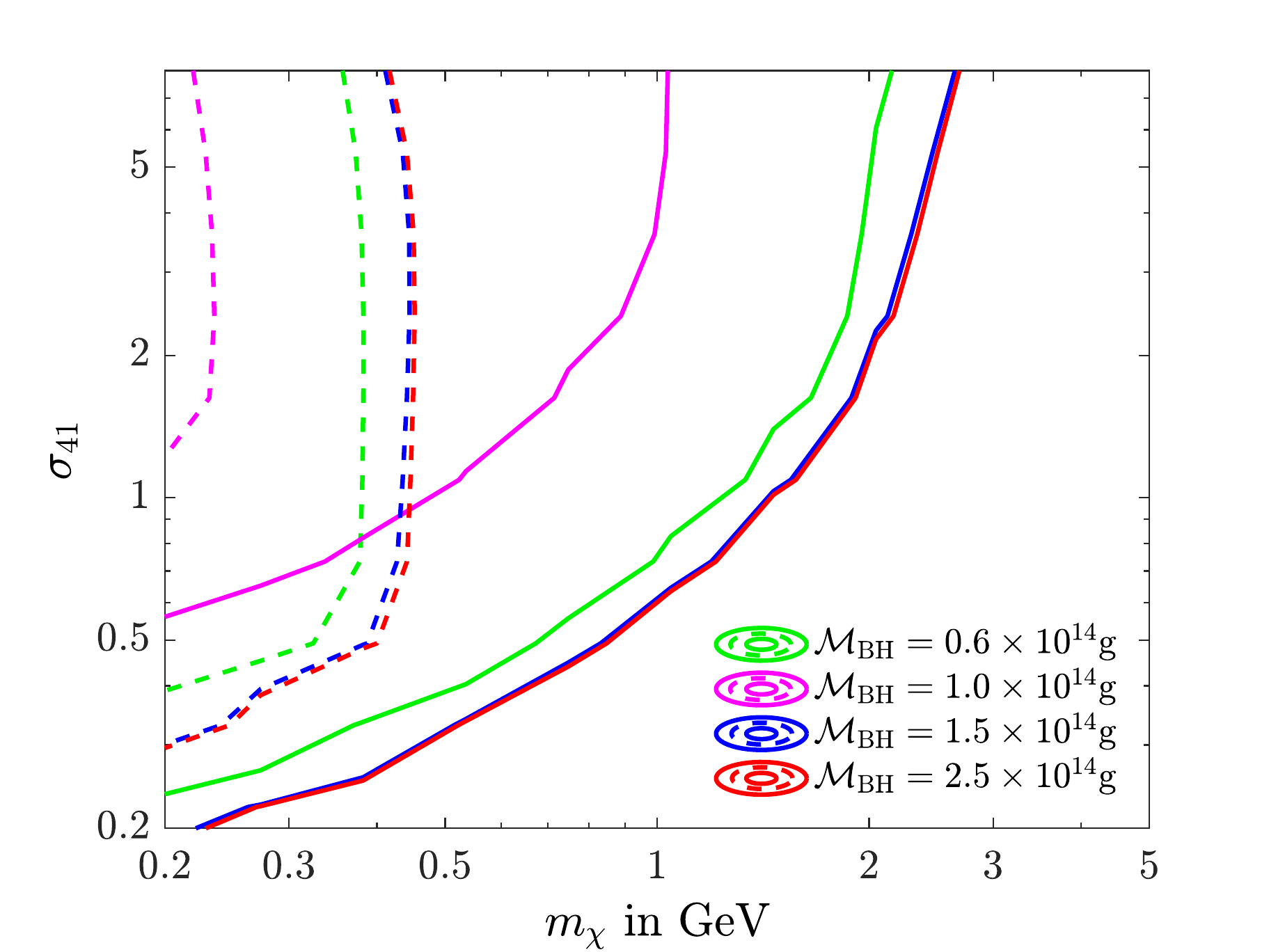}
		\caption{\label{fig:mchi_sigma} The allowed region in the $\sigma_{41}-m_{\chi}$ parameter plane for different PBH masses where, the chosen values of $M_{\rm HDM}$ and $\mathcal{K}$ are the best fit values, obtained from the $\chi^2$ analysis using the IceCube 7.5 year data \cite{IC_7.5yr}. The different coloured solid lines are addressing the bound corresponds to $\Delta T_{21}=-300$ mK for different PBH masses while the dashed lines are representing the same for $\Delta T_{21}=-1000$ mK. Consequently, in each case, the region between the solid and the corresponding dashed line (of individual colour) indicates the allowed zone in the $m_{\chi}$-$\sigma_{41}$ plane.}
	\end{figure}
	We also address bounds in the $m_{\chi}$-$\sigma_{41}$ space for different values of PBH masses ($\mathcal{M_{\rm BH}}$) for the best fit values of $M_{\rm HDM}$ and $\mathcal{K}$ (obtained from $\chi^2$ fit of IceCube results). The results are shown in Fig.~\ref{fig:mchi_sigma}. In Fig.~\ref{fig:mchi_sigma}, the area between the solid and dashed lines (of the same colour) represents the allowed range in the $m_{\chi}$-$\sigma_{41}$ space for particular chosen value of $\mathcal{M_{\rm BH}}$. It can be seen that as the value of $\mathcal{M}_{\rm BH}$ decreases upto around $10^{14}$g, the allowed zone shifts toward lower values of $m_{\chi}$ and higher values of $\sigma_{41}$ (for $\mathcal{M}_{\rm BH}\lessapprox10^{14}$g). But in the case when $\mathcal{M}_{\rm BH}=0.6\times10^{14}$g, the $\sigma_{41}$-$m_{\chi}$ allowed region appears to shift to higher $m_{\chi}$ - lower $\sigma_{41}$ domain. It is shown that this trend in the shift of $m_{\chi}$-$\sigma_{41}$ allowed zone toward the opposite direction sets in for $\mathcal{M}_{\rm BH} \lessapprox 1 \times 10^{14}$ g. For PBH mass $\sim 0.6 \times 10^{14}$ g (\ref{fig:mchi_sigma}), the black holes evaporate very early. Therefore such PBHs are unable to heat up the baryonic fluid significantly. As a consequence, in this particular case, the allowed region is close to those for higher values of $\mathcal{M}_{\rm BH}$. In all four cases, the $\beta_{\rm BH}$ is chosen to be $10^{-29}$.
	It is to be noted that, for the best fit values of $M_{\rm HDM}$ and $\mathcal{K}$, the maximum possible value of the $m_{\chi}$ is $2\sim3$ GeV, when the energy injection by PBH is comparatively low and agree with the results of Barkana \cite{rennan_3GeV} (i.e. $m_{\chi} \leq 3$ GeV).
	
\section{\label{sec:conc} Summary and Discussions}
	In this work, two possible multimessenger signals from the decay of a heavy dark matter are addressed. One is the neutrino signals (of $\sim$ PeV energies) from rare decays of such heavy dark matter at the IceCube neutrino experiment and the other is the influence of this decay process on the absorption temperature of 21-cm Hydrogen signal.
	
	A heavy or super heavy dark matter -- non-thermal particle candidates -- can be produced in the very early Universe, after the inflationary phase during preheating or reheating era or during the GUT phase \cite{PhysRevD.59.023501,Kuzmin1998,PhysRevLett.81.4048,PhysRevD.60.063504,PhysRevD.64.043503,gelmini2010dm,bertone_2010,khlopov_HDM,khlopov_HDM_1}. They may be produced from nonlinear quantum effects, as a consequence of inflaton decays or by gravitational production mechanisms. These long-lived particles can undergo rare decays to produce leptons as the end product. Such decays of super heavy dark matter proceed via QCD parton cascades and these are discussed in Ref.~\cite{kuz,bera,bera1,HILL1983469,PhysRevD.51.4079,PhysRevLett.79.5202,bera2}. The decay cascades are addressed using Dokshitzer-Grivov-Lipatov-Altarelli-Parisi equations or DGLAP equations in short. The electroweak radiative corrections can also be incorporated into the numerical evolution of DGLAP equations. In the hadronic channel of the decay via QCD cascades, the heavy or super heavy dark matter first decays to $q{\bar q}$ which then hadronizes to produce leptons and $\gamma$s as the final products. In this work, the numerical evolutions and Monte Carlo simulations are performed to obtain the spectrum of neutrinos as the end product of the decay cascade of the heavy dark matter of a given mass. We have verified that the neutrino spectrum obtained from the leptonic decay channels via electroweak cascade is not very useful and relevant for the PeV neutrino energy window of IceCube results. Hence in the present calculation only the hadronic channel is included. 
	
	On the other hand, the brightness temperature ($T_{21}$) of 21-cm radio signal of hydrogen depends on the spin temperature $T_s$, the matter temperature or baryon temperature $T_b$ ($T_b$ and $T_s$ are coupled and at some epoch are same) and the background temperature $T_{\gamma}$. The EDGES experiment has measured $T_{21}$ during the era of reionization at the redshift region $14<z<20$ with the central value of $z \sim 17.2$. The $T_{21}$ results along with its $99\%$ C.L. ($-500^{+200}_{-500}$ mK) values appear to be cooler than the value of $T_{21} \sim -200$ mK calculated from the astrophysical and cosmological consideration at the same redshift. With $T_{\gamma}$, the cosmic background radiation temperature, the temperature $T_s$ is influenced or modified by the processes that inject or absorb heat from the system. This can be thought to happen by possible collisions of dark matter with the baryons, the effects of Ly$\alpha$ forest, etc., or other processes through which the baryons can be heated up or cooled down. In this work, we have also included and studied the effects of the evaporations of primordial black holes or PBHs on the evolution of $T_b$. It is found that the mass $m_{\chi}$ of the dark matter that collides with the baryon and affects the baryon temperature (and hence $T_{21}$) should be in the range of hundreds of MeV to a few GeV (around upto 2-3 GeV for PBH mass of $\sim 10^{14}$ g). These dark matter particles are thermally produced and their interaction strength is in the weak interaction regime and are assumed to account for the dark matter relic abundance of the Universe. The fraction of heavy dark matter whose rare decay is addressed in this work related to the multimessenger study is considered to be very small. The multimessenger study is performed with respect to the possible effects on $T_{21}$ of the rare decay of primordial heavy dark matter. To this end, in this work, the heat exchange due to the processes of heavy dark matter decaying into photons ($\gamma$), $e^-$s and $e^+$s are also included in the evolution equation of $T_b$.
	
	With this motivation, the neutrino spectrum from a primordial heavy dark matter decay is first computed and then compared with the IceCube neutrino signal. For this purpose, 7.5 year IceCube up-going muon data are chosen in the neutrino energy window of $2\times 10^5$ GeV to $4\times10^6$ GeV. It contains four data points. The calculated neutrino spectrum is then fitted with the IceCube results and the best fit value of mass and decay width (the product $\mathcal{K}$ of decay width and the density fraction) of the heavy dark matter are obtained. The other possible multimessenger signal from the heavy dark matter decay in relation to its influence on $T_{21}$ signal is addressed by including the effect of this decay in the formalism for computing $T_{21}$ temperature. To this end, the effect of heavy dark matter decay has been included in the evolution equation of baryon temperature. In addition, the evaporation effects of primordial black holes via Hawking radiation and baryon-dark matter (low mass dark matter as mentioned above) collision are also incorporated and $T_{21}$ is computed at various redshift values. The EDGES observational results for $T_{21}$ are then used to constrain the $\mathcal{K} - M_{\rm HDM}$ plane. It is found that both the multimessenger signals considered for heavy dark matter decay signal, namely the IceCube PeV neutrinos and $T_{21}$ temperature are satisfied when the WIMP type dark matter mass $m_{\chi}$ (with which the baryons in the early Universe collide to influence $T_{21}$ signal) is $m_{\chi} \geq 0.4$ GeV and PBH mass $\mathcal{M}_{\rm BH} \approx 10^{14}$ GeV. 
	
	In addition, we also explore how the mass of PBH affects the dark matter-baryon scattering cross-section ($\sigma_{41}$) for different possible dark matter masses ($m_{\chi}$) of few GeV in presence of heavy dark matter decay contributions to $T_{21}$. To this end, we use $T_{21}$ results and constrain $\sigma_{41} - m_{\chi}$ plane. The heavy dark matter mass $M_{\rm HDM}$ and decay width $\mathcal{K}$ are kept fixed at their best fit values ($M_{\rm HDM} = 2.75\times10^8$ GeV, $\mathcal{K} = 2.56\times10^{-29}$ sec$^{-1}$) obtained from the analysis of IceCube 7.5 year HESE data. We find $m_{\chi} < 3$ GeV in order to satisfy the EDGES results for $\mathcal{M}_{\rm BH} \leq 2.5 \times 10^{14}$ g. 
	
	Thus this work is a detailed multimessenger study of the possible heavy dark matter decay signals with IceCube 7.5 year data and 21-cm absorption line temperature of hydrogen and EDGES results, while simultaneously exploring the effects of Ly$\alpha$ forest and PBH evaporation on 21-cm signal.
	
\section*{Acknowledgements}
	At the very outset the authors would like to thank Tracy Slatyer for her suggestions regarding the calculations of energy deposition. Two of the authors (A.H. and R.B.) wish to acknowledge the support received from St. Xavier’s College, Kolkata. A.H. thanks the University Grant Commission (UGC) of the Government of India, for providing financial support. M.P. thanks the DST-INSPIRE fellowship  (DST/INSPIRE/FELLOWSHIP/IF160004) grant by DST, Govt. of India. One of the authors (R.B.) also thanks the Women Scientist Scheme-A fellowship (SR/WOS-A/PM-49/2018), Department of Science and Technology (DST), Govt. of India, for providing financial support. 

\bibliography{PUB21}

\end{document}